\RequirePackage{lineno}
\documentclass[aps,prd,superscriptaddress,showpacs,,amsmath,amssymb,floatfix]{revtex4-1}

\usepackage{subfigure}
\usepackage{graphicx} % Include figure files
\usepackage{dcolumn}  % Align table columns on decimal point

\graphicspath{{ps}}

\begin{document}

%\vspace*{-3\baselineskip}
%\resizebox{!}{3cm}{\includegraphics{belle.eps}}

\preprint{\vbox{ \hbox{   }
                        \hbox{Belle Preprint 2016-07}
                        \hbox{KEK Preprint 2016-07}
%                        \hbox{Belle DRAFT {\it 1380-v11}, July 2016}
%                        \hbox{Intended for {\it Physical Review D}}
%                        \hbox{Author: John Yelton}
%                        \hbox{Committee: Yuji Kato(chair),}
%                        \hbox{Byung Gu Cheon, Bryan Fulsom }
  		              % \hbox{hep-ex nnnn}
}}

\title{ \quad\\[1.0cm] Study of Excited $\Xi_c$ States Decaying into $\Xi_c^0$ and $\Xi_c^+$ Baryons}

%%%% >>>>> insert the authorlist here. BEFORE the abstract !!!!! <<<<<
%%%% >>>>> from the authorship confirmation web page
%%% Name the file author.tex and use \input{author} to insert into your latex file.

%%% Paper:    Excited-Xi_c decays
%%% Journal:  Physical Review D
%%% Contacts: J. Yelton (yelton@phys.ufl.edu)
%%% Non-responding authors or those who said NO are commented out.
%%% ====================================================================
%%% Click the RELOAD button on your web browser to see the updated file.
%%% ====================================================================
%%% Use \input{author} to insert this material into your latex file.
%%%%% Force institutions to appear in alphabetical order when typeset.
\noaffiliation
%%%\affiliation{Aligarh Muslim University, Aligarh 202002}
\affiliation{University of the Basque Country UPV/EHU, 48080 Bilbao}
\affiliation{Beihang University, Beijing 100191}
%%%\affiliation{University of Bonn, 53115 Bonn}
\affiliation{Budker Institute of Nuclear Physics SB RAS, Novosibirsk 630090}
\affiliation{Faculty of Mathematics and Physics, Charles University, 121 16 Prague}
%%%\affiliation{Chiba University, Chiba 263-8522}
\affiliation{Chonnam National University, Kwangju 660-701}
\affiliation{University of Cincinnati, Cincinnati, Ohio 45221}
\affiliation{Deutsches Elektronen--Synchrotron, 22607 Hamburg}
\affiliation{University of Florida, Gainesville, Florida 32611}
%%%\affiliation{Department of Physics, Fu Jen Catholic University, Taipei 24205}
\affiliation{Justus-Liebig-Universit\"at Gie\ss{}en, 35392 Gie\ss{}en}
\affiliation{Gifu University, Gifu 501-1193}
%%%\affiliation{II. Physikalisches Institut, Georg-August-Universit\"at G\"ottingen, 37073 G\"ottingen}
\affiliation{SOKENDAI (The Graduate University for Advanced Studies), Hayama 240-0193}
%%%\affiliation{Gyeongsang National University, Chinju 660-701}
\affiliation{Hanyang University, Seoul 133-791}
\affiliation{University of Hawaii, Honolulu, Hawaii 96822}
\affiliation{High Energy Accelerator Research Organization (KEK), Tsukuba 305-0801}
\affiliation{J-PARC Branch, KEK Theory Center, High Energy Accelerator Research Organization (KEK), Tsukuba 305-0801}
%%%\affiliation{Hiroshima Institute of Technology, Hiroshima 731-5193}
\affiliation{IKERBASQUE, Basque Foundation for Science, 48013 Bilbao}
%%%\affiliation{University of Illinois at Urbana-Champaign, Urbana, Illinois 61801}
%%%\affiliation{Indian Institute of Science Education and Research Mohali, SAS Nagar, 140306}
\affiliation{Indian Institute of Technology Bhubaneswar, Satya Nagar 751007}
\affiliation{Indian Institute of Technology Guwahati, Assam 781039}
\affiliation{Indian Institute of Technology Madras, Chennai 600036}
\affiliation{Indiana University, Bloomington, Indiana 47408}
\affiliation{Institute of High Energy Physics, Chinese Academy of Sciences, Beijing 100049}
\affiliation{Institute of High Energy Physics, Vienna 1050}
\affiliation{Institute for High Energy Physics, Protvino 142281}
%%%\affiliation{Institute of Mathematical Sciences, Chennai 600113}
%%%\affiliation{INFN - Sezione di Torino, 10125 Torino}
\affiliation{Advanced Science Research Center, Japan Atomic Energy Agency, Naka 319-1195}
\affiliation{J. Stefan Institute, 1000 Ljubljana}
\affiliation{Kanagawa University, Yokohama 221-8686}
\affiliation{Institut f\"ur Experimentelle Kernphysik, Karlsruher Institut f\"ur Technologie, 76131 Karlsruhe}
%%%\affiliation{Kavli Institute for the Physics and Mathematics of the Universe (WPI), University of Tokyo, Kashiwa 277-8583}
\affiliation{Kennesaw State University, Kennesaw, Georgia 30144}
\affiliation{King Abdulaziz City for Science and Technology, Riyadh 11442}
%%%\affiliation{Department of Physics, Faculty of Science, King Abdulaziz University, Jeddah 21589}
\affiliation{Korea Institute of Science and Technology Information, Daejeon 305-806}
\affiliation{Korea University, Seoul 136-713}
\affiliation{Kyoto University, Kyoto 606-8502}
\affiliation{Kyungpook National University, Daegu 702-701}
\affiliation{\'Ecole Polytechnique F\'ed\'erale de Lausanne (EPFL), Lausanne 1015}
\affiliation{P.N. Lebedev Physical Institute of the Russian Academy of Sciences, Moscow 119991}
\affiliation{Faculty of Mathematics and Physics, University of Ljubljana, 1000 Ljubljana}
\affiliation{Ludwig Maximilians University, 80539 Munich}
\affiliation{Luther College, Decorah, Iowa 52101}
\affiliation{University of Maribor, 2000 Maribor}
\affiliation{Max-Planck-Institut f\"ur Physik, 80805 M\"unchen}
\affiliation{School of Physics, University of Melbourne, Victoria 3010}
%%%\affiliation{Middle East Technical University, 06531 Ankara}
\affiliation{University of Miyazaki, Miyazaki 889-2192}
\affiliation{Moscow Physical Engineering Institute, Moscow 115409}
\affiliation{Moscow Institute of Physics and Technology, Moscow Region 141700}
\affiliation{Graduate School of Science, Nagoya University, Nagoya 464-8602}
%%%\affiliation{Kobayashi-Maskawa Institute, Nagoya University, Nagoya 464-8602}
%%%\affiliation{Nara University of Education, Nara 630-8528}
\affiliation{Nara Women's University, Nara 630-8506}
\affiliation{National Central University, Chung-li 32054}
\affiliation{National United University, Miao Li 36003}
\affiliation{Department of Physics, National Taiwan University, Taipei 10617}
\affiliation{H. Niewodniczanski Institute of Nuclear Physics, Krakow 31-342}
%%%\affiliation{Nippon Dental University, Niigata 951-8580}
\affiliation{Niigata University, Niigata 950-2181}
\affiliation{University of Nova Gorica, 5000 Nova Gorica}
\affiliation{Novosibirsk State University, Novosibirsk 630090}
%%%\affiliation{Osaka City University, Osaka 558-8585}
%%%\affiliation{Osaka University, Osaka 565-0871}
\affiliation{Pacific Northwest National Laboratory, Richland, Washington 99352}
%%%\affiliation{Panjab University, Chandigarh 160014}
%%%\affiliation{Peking University, Beijing 100871}
\affiliation{University of Pittsburgh, Pittsburgh, Pennsylvania 15260}
%%%\affiliation{Punjab Agricultural University, Ludhiana 141004}
%%%\affiliation{Research Center for Electron Photon Science, Tohoku University, Sendai 980-8578}
%%%\affiliation{Research Center for Nuclear Physics, Osaka University, Osaka 567-0047}
\affiliation{Theoretical Research Division, Nishina Center, RIKEN, Saitama 351-0198}
%%%\affiliation{RIKEN BNL Research Center, Upton, New York 11973}
%%%\affiliation{Saga University, Saga 840-8502}
\affiliation{University of Science and Technology of China, Hefei 230026}
%%%\affiliation{Seoul National University, Seoul 151-742}
%%%\affiliation{Shinshu University, Nagano 390-8621}
\affiliation{Showa Pharmaceutical University, Tokyo 194-8543}
\affiliation{Soongsil University, Seoul 156-743}
%%%\affiliation{University of South Carolina, Columbia, South Carolina 29208}
\affiliation{Stefan Meyer Institute for Subatomic Physics, Vienna 1090}
\affiliation{Sungkyunkwan University, Suwon 440-746}
\affiliation{School of Physics, University of Sydney, New South Wales 2006}
\affiliation{Department of Physics, Faculty of Science, University of Tabuk, Tabuk 71451}
\affiliation{Tata Institute of Fundamental Research, Mumbai 400005}
\affiliation{Excellence Cluster Universe, Technische Universit\"at M\"unchen, 85748 Garching}
\affiliation{Department of Physics, Technische Universit\"at M\"unchen, 85748 Garching}
\affiliation{Toho University, Funabashi 274-8510}
%%%\affiliation{Tohoku Gakuin University, Tagajo 985-8537}
\affiliation{Department of Physics, Tohoku University, Sendai 980-8578}
\affiliation{Earthquake Research Institute, University of Tokyo, Tokyo 113-0032}
\affiliation{Department of Physics, University of Tokyo, Tokyo 113-0033}
\affiliation{Tokyo Institute of Technology, Tokyo 152-8550}
\affiliation{Tokyo Metropolitan University, Tokyo 192-0397}
%%%\affiliation{Tokyo University of Agriculture and Technology, Tokyo 184-8588}
%%%\affiliation{University of Torino, 10124 Torino}
%%%\affiliation{Toyama National College of Maritime Technology, Toyama 933-0293}
%%%\affiliation{Utkal University, Bhubaneswar 751004}
\affiliation{Virginia Polytechnic Institute and State University, Blacksburg, Virginia 24061}
\affiliation{Wayne State University, Detroit, Michigan 48202}
\affiliation{Yamagata University, Yamagata 990-8560}
\affiliation{Yonsei University, Seoul 120-749}
% \author{A.~Abdesselam}\affiliation{Department of Physics, Faculty of Science, University of Tabuk, Tabuk 71451} % Tabuk
  \author{J.~Yelton}\affiliation{University of Florida, Gainesville, Florida 32611} % Florida
  \author{I.~Adachi}\affiliation{High Energy Accelerator Research Organization (KEK), Tsukuba 305-0801}\affiliation{SOKENDAI (The Graduate University for Advanced Studies), Hayama 240-0193} % KEK
% \author{K.~Adamczyk}\affiliation{H. Niewodniczanski Institute of Nuclear Physics, Krakow 31-342} % Krakow
  \author{H.~Aihara}\affiliation{Department of Physics, University of Tokyo, Tokyo 113-0033} % Tokyo
% \author{S.~Al~Said}\affiliation{Department of Physics, Faculty of Science, University of Tabuk, Tabuk 71451}\affiliation{Department of Physics, Faculty of Science, King Abdulaziz University, Jeddah 21589} % Tabuk
% \author{K.~Arinstein}\affiliation{Budker Institute of Nuclear Physics SB RAS, Novosibirsk 630090}\affiliation{Novosibirsk State University, Novosibirsk 630090} % BINP
% \author{Y.~Arita}\affiliation{Graduate School of Science, Nagoya University, Nagoya 464-8602} % Nagoya
\author{D.~M.~Asner}\affiliation{Pacific Northwest National Laboratory, Richland, Washington 99352} % PNNL
% \author{T.~Aso}\affiliation{Toyama National College of Maritime Technology, Toyama 933-0293} % Toyama
% \author{H.~Atmacan}\affiliation{Middle East Technical University, 06531 Ankara} % METU
  \author{V.~Aulchenko}\affiliation{Budker Institute of Nuclear Physics SB RAS, Novosibirsk 630090}\affiliation{Novosibirsk State University, Novosibirsk 630090} % BINP
  \author{T.~Aushev}\affiliation{Moscow Institute of Physics and Technology, Moscow Region 141700} % MIPT
  \author{R.~Ayad}\affiliation{Department of Physics, Faculty of Science, University of Tabuk, Tabuk 71451} % Tabuk
% \author{T.~Aziz}\affiliation{Tata Institute of Fundamental Research, Mumbai 400005} % Tata
% \author{V.~Babu}\affiliation{Tata Institute of Fundamental Research, Mumbai 400005} % Tata
  \author{I.~Badhrees}\affiliation{Department of Physics, Faculty of Science, University of Tabuk, Tabuk 71451}\affiliation{King Abdulaziz City for Science and Technology, Riyadh 11442} % Tabuk
  \author{S.~Bahinipati}\affiliation{Indian Institute of Technology Bhubaneswar, Satya Nagar 751007} % IITB
  \author{A.~M.~Bakich}\affiliation{School of Physics, University of Sydney, New South Wales 2006} % Sydney
% \author{A.~Bala}\affiliation{Panjab University, Chandigarh 160014} % Panjab
% \author{Y.~Ban}\affiliation{Peking University, Beijing 100871} % Peking
% \author{V.~Bansal}\affiliation{Pacific Northwest National Laboratory, Richland, Washington 99352} % PNNL
% ...
% Ljubljana
% \author{N.~Zwahlen}\affiliation{\'Ecole Polytechnique F\'ed\'erale de Lausanne (EPFL), Lausanne 1015} % Lausanne
% \author{O.~Zyukova}\affiliation{Budker Institute of Nuclear Physics SB RAS, Novosibirsk 630090}\affiliation{Novosibirsk State University, Novosibirsk 630090} % BINP
\collaboration{The Belle Collaboration}

%\author{Author}\affiliation{affiliation}
%\collaboration{The Belle Collaboration}
%\noaffiliation
%% end author list

\begin{abstract}
Using a data sample of 980 ${\rm fb}^{-1}$ of $e^+e^-$ annihilation data taken with the Belle detector
operating at the KEKB asymmetric-energy $e^+e^-$ collider, we report the results of a study of excited $\Xi_c$ states
that decay, via the emission of photons and/or charged pions, into $\Xi_c^0$ or $\Xi_c^+$ ground state 
charmed-strange baryons. 
We present new measurements of the masses of all members of the $\Xi_c^{\prime}$, $\Xi_c(2645)$, $\Xi_c(2790)$,
$\Xi_c(2815)$, and $\Xi_c(2980)$ isodoublets, measurements of the intrinsic widths of those that decay
strongly, and evidence of previously unknown transitions.

\end{abstract}


\maketitle

%%%% >>>> keep the final version single-spaced

{\renewcommand{\thefootnote}{\fnsymbol{footnote}}}
\setcounter{footnote}{0}

\section*{Introduction}
The $\Xi_c$ states consist of a combination of a charm quark, a strange quark, and an up or down quark~\cite{CC}.
The ground-state $\Xi_c^0$ and $\Xi_c^+$ have $J^P$ = $1/2^+$ and no orbital angular momentum 
and, like the $\Lambda_c^+$, have a wave-function that is antisymmetric under interchange of the lighter 
quark flavors or spins. 
The two ground states are the only members of the group that decay weakly, and their masses, lifetimes, and many of their 
decay modes have been measured~\cite{PDG}. The $\Xi_c$ states also exist in many angular momentum configurations 
of the constituent quarks, each as an isospin pair. These excited states have been found to decay either electromagnetically or 
strongly in three different general types of decay:
to the $\Xi_c$ ground states
together with mesons and/or photons, to final states that include a $\Lambda_c^+$ and a kaon~\cite{CHISTOV,AUBERT,KATO2}, 
and to 
$\Lambda D$ final states~\cite{Kato}.
This paper concentrates on the measurements of the masses and widths of the five isospin pairs of excited $\Xi_c$ baryons
that include a ground-state $\Xi_c$ in their decay chain. All five pairs under investigation have previously been discovered, 
but in general their masses and intrinsic widths have not been measured precisely. 

In the Heavy Quark Symmetry (HQS)~\cite{HQS} picture, the $\Xi_c$ baryons are the combination of a heavy (charm) quark 
and a light ($us$ or $ds$) di-quark. In the standard quark model, the first excited pair have, like the ground-states,  $J^P$ = $1/2^+$, 
but in this case the quarks in the light di-quark are symmetric under interchange
in a similar manner to the $\Sigma$ and $\Sigma_c$ baryons. These are known as $\Xi_c^{\prime}$ baryons. This isospin pair
is the only one that, because of the small mass differences involved, decays electromagnetically. The $\Xi_c^{\prime}$ pair was first
discovered by CLEO in 1999~\cite{Xicp}, but there have been no high statistics measurements of its mass. When the same ``symmetric''
di-quark combines with the charm quark in a $J^P$ = $3/2^+$ configuration, the particles are denoted as $\Xi_c(2645)$, and sometimes
known as $\Xi_c^*$.  These were first
discovered by CLEO in 1995~\cite{Xics} and 1996~\cite{Xics2}, and subsequently confirmed by a series of other experiments.   

In the HQS model of baryons, the first orbitally excited states have a unit of angular momentum between the heavy quark and the light di-quark, with 
a spin-zero configuration of the light di-quark. This yields a total spin associated with the light degrees of freedom of one unit. 
This one unit then combines with the heavy quark spin to make  $J^P$ = $1/2^-$ and $J^P$ = $3/2^-$ isodoublets. 
These two isodoublets were discovered by 
CLEO~\cite{half,threehalves}; the lower mass ($\Xi_c(2790)$, with $J^P$ = $1/2^-$) states were
found in the mode $\Xi_c^{\prime}\pi$ and the higher mass 
($\Xi_c(2815)$, with $J^P$ = $3/2^-$) states
in $\Xi_c(2645)\pi$ combinations. Until now, no measurements of their intrinsic widths have been reported. 
The quark configurations of all these states 
have been identified purely by their masses and decay products, but they fit the expected patterns so well that 
their identification is not considered controversial. 
A fifth isodoublet of excited states, the $\Xi_c(2980)$, was 
discovered by Belle~\cite{CHISTOV} and confirmed by BaBar~\cite{AUBERT} in the decay mode $\Lambda_c^+K^-\pi^+$,
and subsequently seen, also by Belle, in the mode $\Xi_c(2645)\pi$~\cite{LESIAK}. Its $J^P$ value is not determined, nor can it be assigned
trivially in the standard quark model.
Higher mass $\Xi_c$ excited states of unknown quark configuration have been found~\cite{CHISTOV,AUBERT,KATO2}.

The aim of this analysis is to measure, with greater precision than before, the masses of the 
five isodoublets of excited 
$\Xi_c$ baryons that decay with $\Xi_c$ ground states in their decay products, to measure the 
intrinsic widths of the four of these five isodoublets that 
decay strongly, and to look for new transitions that can help the identification of these states. 
Knowing the masses of these particles more accurately is of both
practical and theoretical interest, and measuring their widths can then lead 
to measurements of the matrix elements of their decays; by
HQS, these matrix elements are also applicable to other excited charm and bottom baryons~\cite{HQS,CC2}. 

This analysis uses a data sample of $e^+e^-$ annihilations recorded by the Belle detector~\cite{Belle} operating at the KEKB asymmetric-energy $e^+e^-$
collider~\cite{KEKB}. It corresponds to an integrated luminosity of 980 ${\rm fb}^{-1}$.
The majority of these data were taken with the accelerator energy tuned for production of the $\Upsilon(4S)$ resonance, as this is optimum
for investigation of $B$ decays.
However, the $\Xi_c$ particles in this analysis are produced in continuum charm production and are of 
higher momentum than those that are 
decay products of $B$ mesons, so
the dataset used in this analysis also includes the Belle data taken at beam energies corresponding to the other $\Upsilon$ resonances
and the continuum nearby. 

\section*{The Belle Detector and Ground-State $\Xi_c$ Reconstruction}

The Belle detector is a large solid-angle spectrometer comprising six sub-detectors: the Silicon Vertex Detector (SVD), the 50-layer Central
Drift Chamber (CDC), the Aerogel Cherenkov Counter (ACC), the Time-of-Flight scintillation counter (TOF), 
the electromagnetic calorimeter, and the 
$K_L$ and muon detector. A superconducting solenoid produces a 1.5 T magnetic field throughout the first five of these sub-detectors.
The detector is described in detail elsewhere~\cite{Belle}. Two inner detector configurations were used. The first comprised a 2.0 cm radius beampipe and a 3-layer silicon vertex detector, and the second a 1.5 cm radius beampipe and a 4-layer silicon detector and a small-cell inner drift chamber.

In order to study $\Xi_c$ baryons, we first reconstruct a large sample of ground-state $\Xi_c^0$ and $\Xi_c^+$ 
baryons with good signal-to-noise ratio. To obtain large statistics, we use many decay modes of the ground-states, 
each with specific requirements on their
decay products designed to suppress combinatorial backgrounds. 
The decay modes used are listed in Table~\ref{tab:yields}.

Final-state charged particles, 
$\pi^{\pm}, K^{-}$, and $p$, are selected using the likelihood 
information from the tracking (SVD, CDC) and charged-hadron identification (CDC, ACC, TOF) systems into a combined likelihood, 
${\cal L}(h1:h2) = {\cal L}_{h1}/({\cal L}_{h1} + {\cal L}_{h2})$ 
where $h_1$ and $h_2$ are $p$, $K$, and $\pi$ as appropriate. In general, we require proton candidates 
to have ${\cal L}(p:K)>0.6$ and ${\cal L}(p:\pi)>0.6$, 
kaon candidates to have ${\cal L}(K:p)>0.6$ and ${\cal L}(K:\pi)>0.6$, 
and pions to have the less restrictive requirements of
${\cal L}(\pi:K)>0.2$ and ${\cal L}(\pi:p)>0.2$. 
For one mode, $\Xi_c^0\to pK^-K^-\pi^+$, the proton level is increased to 0.97 to ensure no
contamination from charmed mesons. The $\pi^0$ candidates are reconstructed from two detected neutral clusters in the ECL
each consistent with being
due to a photon and each with an energy greater than 50 ${\rm MeV}$ in the laboratory frame. The invariant mass of the photon pair is 
required to be
within 3 standard deviations ($\sigma$) of the nominal $\pi^0$ mass.
The same sample of photons is used for the
$\Sigma^0\to\Lambda\gamma$ reconstruction.
The $\Lambda\ (K_S^0)$ candidates are made from 
$p\pi\ (\pi\pi)$ pairs with a production vertex significantly
separated from the
interaction point (IP). For the case of the proton from the $\Lambda$, the particle identification is loosened to 
${\cal L}(p:K)>0.2$ and ${\cal L}(p:\pi)>0.2$.  
The $\Lambda$ candidates used to directly reconstruct $\Xi_c$ candidates are required to have an origin consistent with 
the IP, 
but those that are daughters of $\Xi^-$, $\Xi^0$ or $\Omega^-$ candidates are not subject to this
 requirement.

The $\Xi^-$ and $\Omega^-$ candidates are made from the $\Lambda$ candidates detailed above, 
together with a $\pi^-$
or $K^-$ candidate. The vertex formed from the $\Lambda$ and $\pi/K$ is required to be between the IP and the $\Lambda$ 
decay vertex. 

The $\Xi^0$ and $\Sigma^+$ reconstruction is complicated by the fact that the parent hyperon decays with a $\pi^0$ (which has
negligible vertex position information) as one of its daughters. In the case of the $\Sigma^+ \to p \pi^0$ reconstruction, combinations
of $\pi^0$ candidates and protons are made using those protons with a
significantly large ($>1$ mm) impact parameter with respect to the IP. Then, assuming the IP to be the origin, a $\Sigma^+$
trajectory is assumed, and the intersection of this trajectory with the reconstructed proton trajectory found. 
This position is taken as the decay location of the $\Sigma^+$ hyperon, and the $\pi^0$ is then re-fit using this as its point of 
origin. Only those combinations with the decay location of the $\Sigma^+$ indicating a positive $\Sigma^+$ path length are retained.
The $\Xi^0$ is reconstructed along similar lines but, in this case, it is not necessary to require a large
impact parameter with respect to the IP.

Mass requirements are placed on all the hyperons reconstructed, based on the canonical masses of these particles. 
The half-widths of the mass intervals, all of which correspond to approximately two standard deviations of the resolution,  are (in ${\rm MeV}/c^2$), 
8.0, 5.0, 3.5, 3.5, 8.0 and 3.5
for $\Sigma^+$, $\Xi^0$, $\Xi^-$, $\Omega^-$, $\Sigma^0$, and $\Lambda$, 
respectively; the particles 
are kinematically constrained to the nominal masses for further analysis.

\begin{figure}[htb]                                                                                                                   
\includegraphics[width=7in]{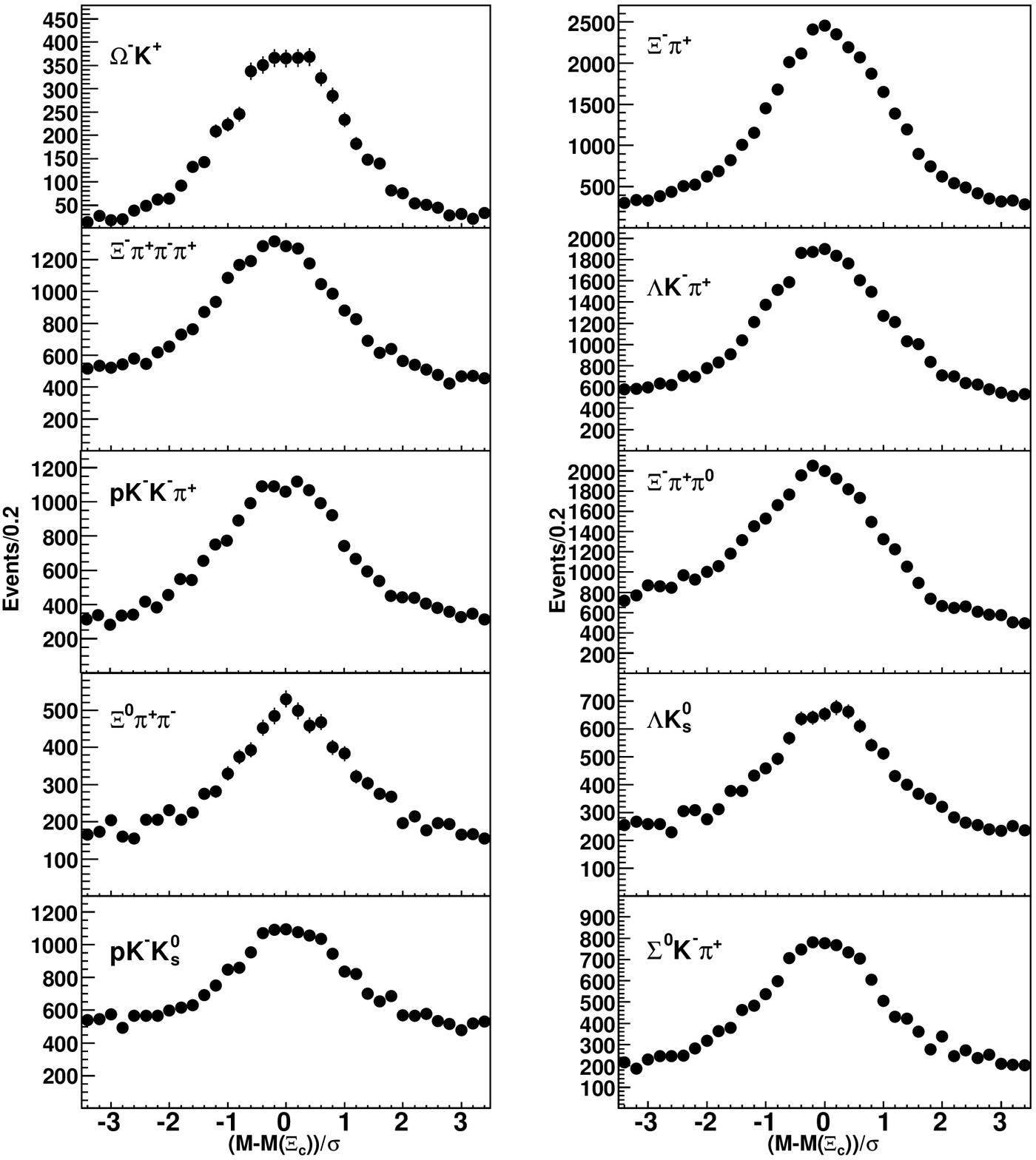}
\caption{ The distribution of the ``pull mass,'' \emph{i.e.} $(M_{\rm measured}-M(\Xi_c^0))/\sigma$, for the ten $\Xi_c^0$ modes 
used in this analysis. 
There is an $x_p > 0.6$ requirement 
applied for presentation purposes, but this is not part of the analysis chain.}

\label{fig:Figure1a}
\end{figure}

\begin{figure}[htb]                                                                                                                   
\includegraphics[width=7in]{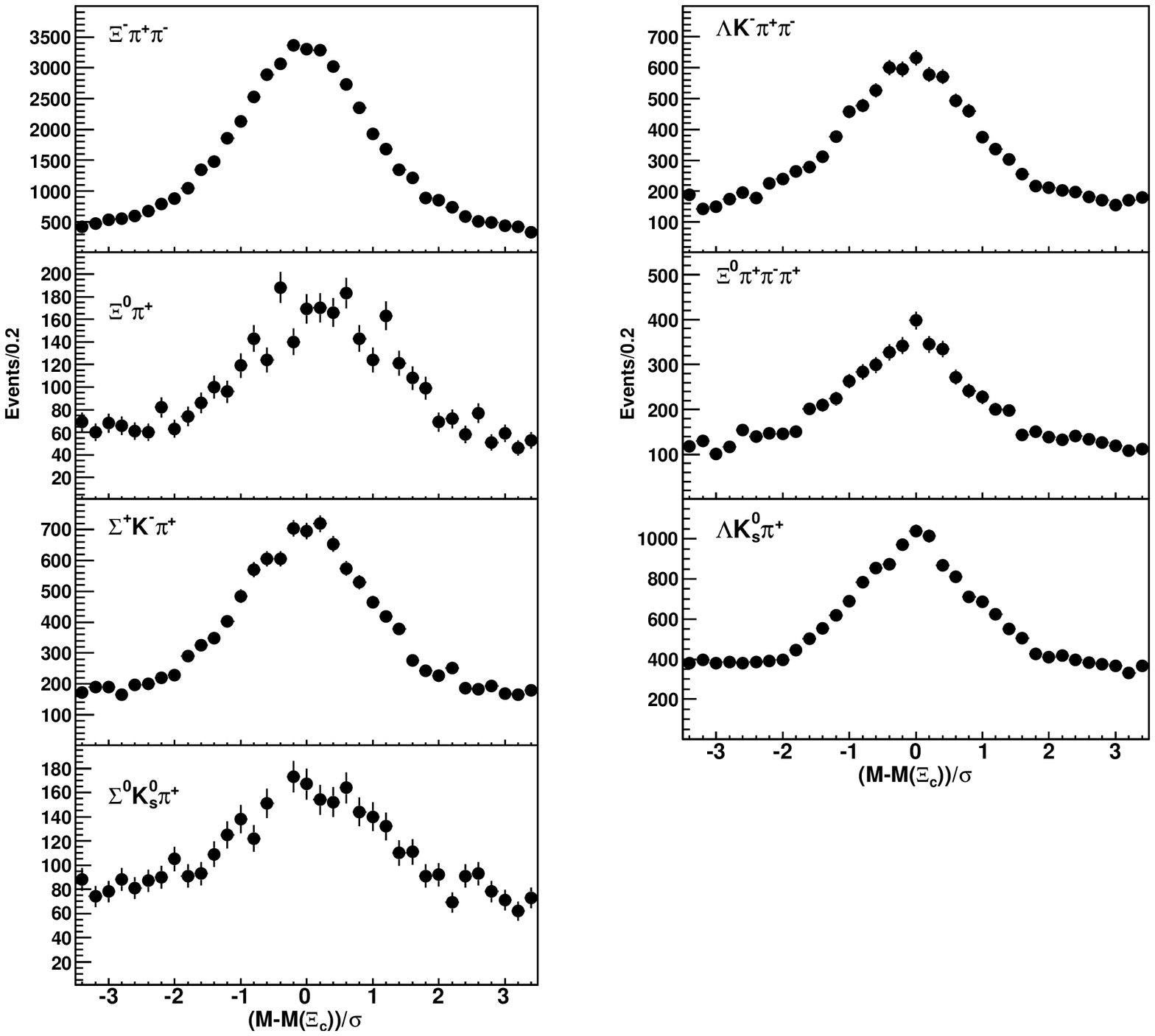}
\caption{ The distribution of the ``pull mass,'' \emph{i.e.} $(M_{\rm measured}-M(\Xi_c^+))/\sigma$, for the seven $\Xi_c^+$ modes 
used in this analysis. 
There is an $x_p > 0.6$ requirement 
applied for presentation purposes, but this is not part of the analysis chain.}
\label{fig:Figure2a}
\end{figure}

To show the yield of the reconstructed $\Xi_c^0$ and $\Xi_c^+$ baryons, we plot the mass peaks in Figs.~\ref{fig:Figure1a} and \ref{fig:Figure2a} 
as distributions of the ``pull mass,'' 
that is, the difference between the measured and nominal mass (2470.85 ${\rm MeV}/c^2$ and 2467.93 
${\rm MeV}/c^2$ for the $\Xi_c^0$ and $\Xi_c^+$, respectively\cite{PDG}), divided by the resolution. For these plots, we require the scaled momentum
of $x_p > 0.6$, where $x_p = p^*c/\sqrt{(s/4 - c^2m^2)}$, and $p^*$ is the momentum in the center of mass and $s$ is the total 
center-of-mass energy squared. 
This cut is not part of the analysis chain as we prefer to place an $x_p$ cut on the excited states; however, it serves to display
the signal-to-noise ratio of our reconstructed ground-state baryons. 
We also show in Table~\ref{tab:yields} the signal and background yields in each decay mode, integrated in a 
$\pm 2 \sigma$ interval around the nominal mass. 
%In all cases in this analysis, we
%use the PDG~\cite{PDG} fit values of 2470.85 $\rm{MeV}/c^2$ and 2467.93 $\rm{MeV}/c^2$ for the $\Xi_c^0$ and $\Xi_c^+$ masses, resp/%ectively. 

Once the daughter particles of a $\Xi_c$ candidate are selected, the $\Xi_c$ candidate itself is made by kinematically fitting the 
daughters to a common decay vertex. The IP is not included in this vertex, as the small decay length associated 
with the $\Xi_c$
decays, though very short compared with the $\Xi^-$, $\Xi^0$, $\Omega^-$ and $\Sigma^+$ decay lengths, is not completely negligible. The 
$\chi^2$ of this vertex is required to be consistent with all the daughters being produced by a common parent.

The above reconstruction of $\Xi_c$ baryons is optimized to have generally 
good efficiency and good signal-to-noise ratio for high momentum candidates.
The identical sample is used for all the different analyses described below.

\begin{table}[htb]
%\caption{ The $\Xi_c$ decay modes used in this analysis}
\caption{The yield of each decay mode of the ground state $\Xi_c^0$ and $\Xi_c^+$ for $x_p >0.6$. The
yields and background are found by integrating over a range of $\pm 2 \sigma$ around the peak value, where $\sigma$ is the resolution. Note that this $x_p$ cut
is not part of the analysis.}

\begin{tabular}
%{@{\hspace{0.5cm}}l@{\hspace{0.5cm}}||@{\hspace{0.5cm}}c@{\hspace{0.5cm}}}
 {@{\hspace{0.5cm}}l@{\hspace{0.5cm}}||  @{\hspace{0.5cm}}c@{\hspace{0.5cm}}|@{\hspace{0.5cm}}c@{\hspace{0.5cm}} }

\hline \hline
Mode & Signal yield $(10^3)$  & Background yield $(10^3)$\\
\hline

$\Omega^-K^+$ & 4.3  & 0.4           \\
$\Xi^-\pi^+$ & 24.3& 6.5          \\
$\Xi^-\pi^+\pi^-\pi^+$ & 9.6 & 9.8 \\
$\Lambda K^-\pi^+$ & 15.7 & 11.3   \\
$pK^-K^-\pi^+$ & 9.5 & 6.5         \\
$\Xi^-\pi^+\pi^0$ & 15.8 & 13.2    \\
$\Xi^0\pi^+\pi^+$ & 3.7 & 3.4      \\
$\Lambda K^0_S$ & 4.8 & 5.0        \\
$pK^-K^0_S$ & 6.4 & 10.6        \\
$\Sigma^0K^-\pi^+$ & 6.7 & 4.3     \\

\hline

 Sum of above $\Xi_c^0$ modes & 100.8  & 71.0  \\
\hline
$\Xi^-\pi^+\pi^+$ & 33.6 & 8.8 \\
$\Lambda K^-\pi^+\pi^+$ & 5.0 & 3.4 \\
$\Xi^0\pi^+$ & 1.4 & 1.1 \\
$\Xi^0\pi^+\pi^-\pi^+$ & 2.5 & 2.4 \\
$\Sigma^+K^-\pi^+$ & 6.0 & 3.5 \\
$\Lambda K^0_S \pi^+$ & 6.5 & 7.4 \\
$\Sigma^0 K^0_S \pi^+$ & 1.1 & 1.5 \\
\hline 
Sum of above $\Xi_c^+$ modes & 56.1 & 28.1 \\
\hline
\hline

\end{tabular}

\label{tab:yields}
\end{table}

\section*{The \boldmath${\Xi_c(2645)}$ and \boldmath${\Xi_c(2815)}$ isodoublets}

\begin{figure}[htb]                                                                                                                                      
\includegraphics[width=7in]{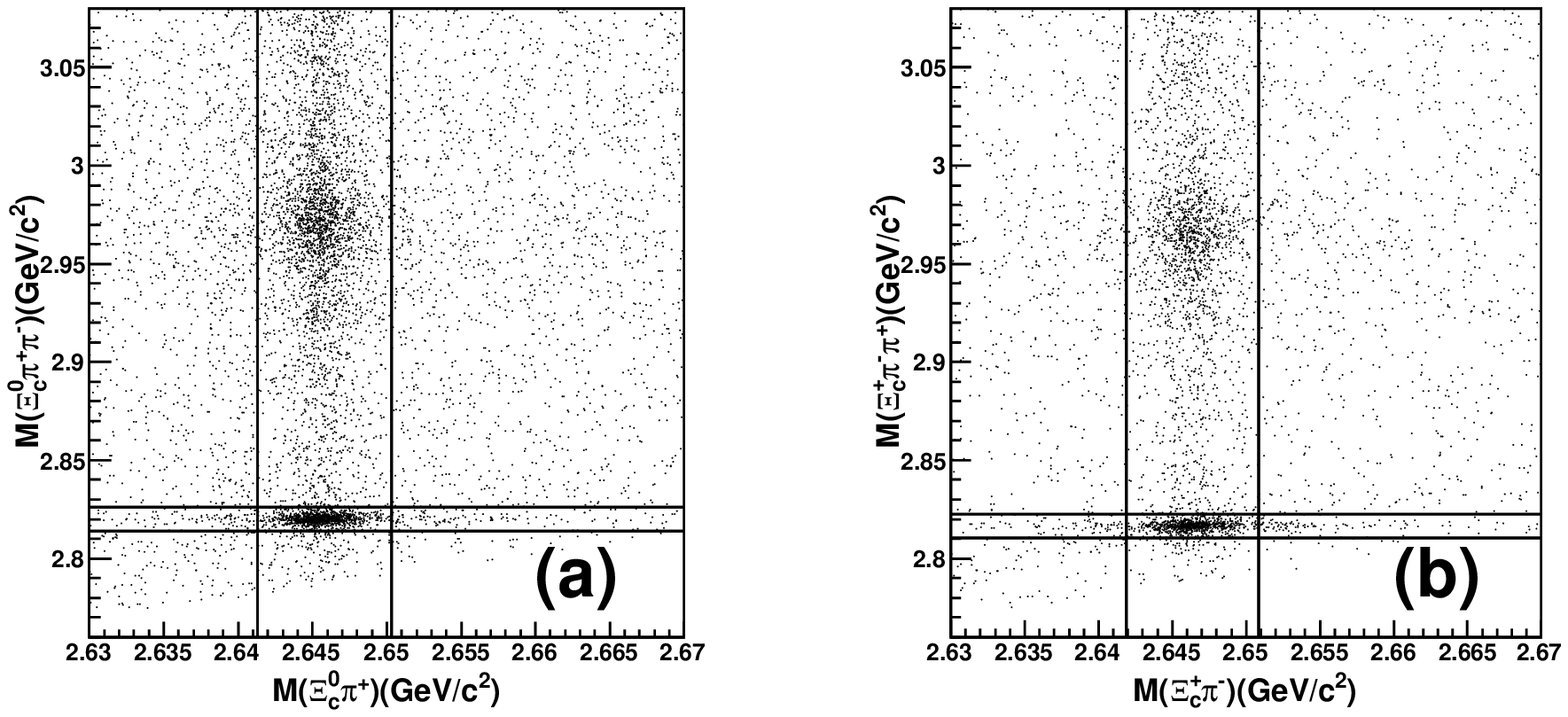}
\caption{ Scatter plots of  M($\Xi_c\pi\pi$) versus  M($\Xi_c(2645))$ for a) $\Xi_c^0$ and b) $\Xi_c^+$.  The lines show the positions of the requirements around the mass
peaks used for projections on to the other axis.}
\label{fig:JAllData28192D}
\end{figure}

The decay chain $\Xi_c(2815)\to\Xi_c(2645)\pi,\ \Xi_c(2645)\to\Xi_c\pi$ allows us to obtain samples of both 
the $\Xi_c(2815)$ and $\Xi_c(2645)$ states 
with excellent signal-to-noise ratio. 
The mass-constrained $\Xi_c$ samples obtained as described above are combined with two appropriately charged pions not 
contributing to the reconstructed $\Xi_c$; a vertex
constraint of the three particles is made with the IP included to optimize the mass resolution. For each isospin state, all
decay modes of the $\Xi_c$ are considered together. 
We then place a requirement of $x_p > 0.7$ on the $\Xi_c(2815)$ candidate. 
This requirement is typical for studies of orbitally excited charmed baryons which are known to have hard fragmentation functions.
The scatter plots of $\Xi_c\pi\pi$ mass versus $\Xi_c\pi$ mass show the significant
 $\Xi_c(2815)\to\Xi_c(2645)\pi$ signals in both isospin states (Fig.~\ref{fig:JAllData28192D}).

\begin{figure}[htb]                                                                                                                                      
\includegraphics[width=7in]{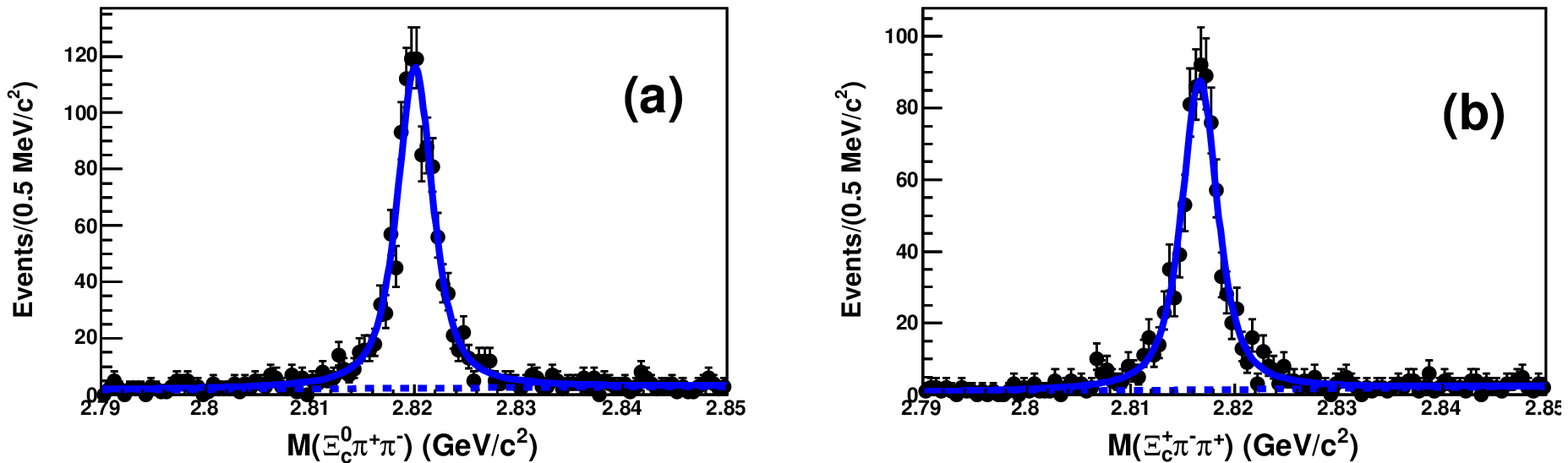}
\caption{ The a) $\Xi_c^0\pi\pi$ and b) $\Xi_c^+\pi\pi$ invariant mass distributions with a cut on $\Xi_c(2645)$ invariant mass of the intermediate state and a scaled momentum requirement of $x_p > 0.7$. Both show significant $\Xi_c(2815) $ signals. 
The fits are described in the text. The dashed lines represent the combinatorial
background contributions. }
\label{fig:JAllData2819}
\end{figure}

To study the $\Xi_c(2815)$ we place a requirement that the $\Xi_c^0\pi^+$ ($\Xi_c^+\pi^-$) combination be within $\pm 5.0$ ${\rm MeV}/c^2$ of the 
$\Xi_c(2645)$ mass peak (as shown by the lines in Fig.~\ref{fig:JAllData28192D}), and then make plots of the
$\Xi_c^0\pi^+\pi^-$ and $\Xi_c^+\pi^-\pi^+$ mass (Fig.~\ref{fig:JAllData2819}). 
In each case, a prominent peak is found in the expected region, with a low level of background. 

These two distributions are fit to signal functions comprising Breit-Wigner functions convolved with double-Gaussian resolution functions 
and first-order polynomial backgrounds. For this and all other distributions in this analysis, 
the resolution function is obtained from Monte Carlo (MC) events, 
generated using EvtGen~\cite{EVTGEN} with the Belle detector response simulated using the GEANT3~\cite{GEANT3} framework. 
The widths of the resolution functions are expressed as weighted averages of the components of the double Gaussian, 
and are shown in Table~\ref{tab:MC}.
The fitted masses are $(2819.98\pm 0.08)$ ${\rm MeV}/c^2$ and $(2816.62\pm0.09)$ ${\rm MeV}/c^2$ 
for the $\Xi_c^+(2815)$ and $\Xi_c^0(2815)$, respectively, 
and the widths are $\Gamma(\Xi_c^+(2815))= (2.34\pm0.18)$ ${\rm MeV}$ and $\Gamma(\Xi_c^0(2815)) = (2.31\pm0.20) $ ${\rm MeV}$; 
in all cases,
the quoted uncertainty is only statistical.  

The Belle momentum scale has been studied using the masses of well-measured parent particles as reference points. 
Close inspection shows small biases in the momentum
measurement of low momentum tracks which can lead to a small mis-measurement of mass peaks found using charged pion transitions
and systematic shifts in the measured
mass of the excited states under investigation here. 

%High-statistics checks were made on the $M(D^{*+})-M(D^0)$ mass difference measured in 
%$D^{*+}\to D^0\pi^+$ decay as a function of the momentum of the transition pion. As have been presented elsewhere~\cite{SIGC}, these studies 
%show a deviation from the expected mass difference which is attributed to 
High-statistic studies of the $M(D^{*+})-M(D^0)$ mass difference measured in $D^{*+}\to D^0\pi^+$ decay as a 
function of the momentum of the transition pion~\cite{SIGC} show deviations from the expected mass difference that are attributed to
limitations in the accuracy with which the track-fitting programs 
take into account the 
detector material. These limitations are reproduced in the Monte Carlo modeling. In the analyses described here, these small biases are
thus taken into account by assigning a correction of a fraction of an ${\rm MeV}/c^2$ 
to the measured masses of all the strongly decaying resonances
under consideration, using the Monte Carlo
programs to evaluate these corrections. 
These corrections have already been applied in the masses for the $\Xi_c(2815)$ states quoted above, and are listed in Table~\ref{tab:MC}.
The systematic uncertainties assigned to these corrections are discussed further below and 
final results including systematic uncertainties shown in Table~\ref{tab:results}.

\begin{table}[htb]

\caption{The width of the Monte Carlo derived resolution functions, expressed as a weighted average in quadrature of the two standard deviations that
comprise the double-Gaussian functions used, and the mass offsets derived from Monte Carlo.}
\label{tab:MC}

\begin{tabular}
 {@{\hspace{0.5cm}}l@{\hspace{0.5cm}}||  @{\hspace{0.5cm}}c@{\hspace{0.5cm}}|@{\hspace{0.5cm}}c@{\hspace{0.5cm}} }
\hline \hline
Mode & Resolution ($\sigma_{\rm av}$(${\rm MeV}/c^2$))	& Monte Carlo $M_{\rm rec}-M_{\rm gen}$(${\rm MeV}/c^2$) \\

\hline

$\Xi_c(2645)\to \Xi_c\pi$    &	$0.82$ &  $-0.07$ \\
$\Xi_c(2815)\to \Xi_c\pi\pi$	&   $1.15$ & $-0.12$ \\
$\Xi_c(2980)\to \Xi_c\pi\pi$	&   $1.99$ & $-0.28$ \\
$\Xi_c^{\prime} \to \Xi_c\gamma$    &  $5.5$ & $+0.36$ \\
$\Xi_c(2790) \to \Xi_c^{\prime}\pi$    &  $1.34$ &  $-0.12$ \\
$\Xi_c(2815) \to \Xi_c^{\prime}\pi$    &  $1.58$ &  $-0.17$ \\
$\Xi_c(2980) \to \Xi_c^{\prime}\pi$    &  $1.90$  & $-0.23$ \\

\hline
\hline

\end{tabular}

\end{table}

\begin{figure}[htb]																	 
\includegraphics[width=7in]{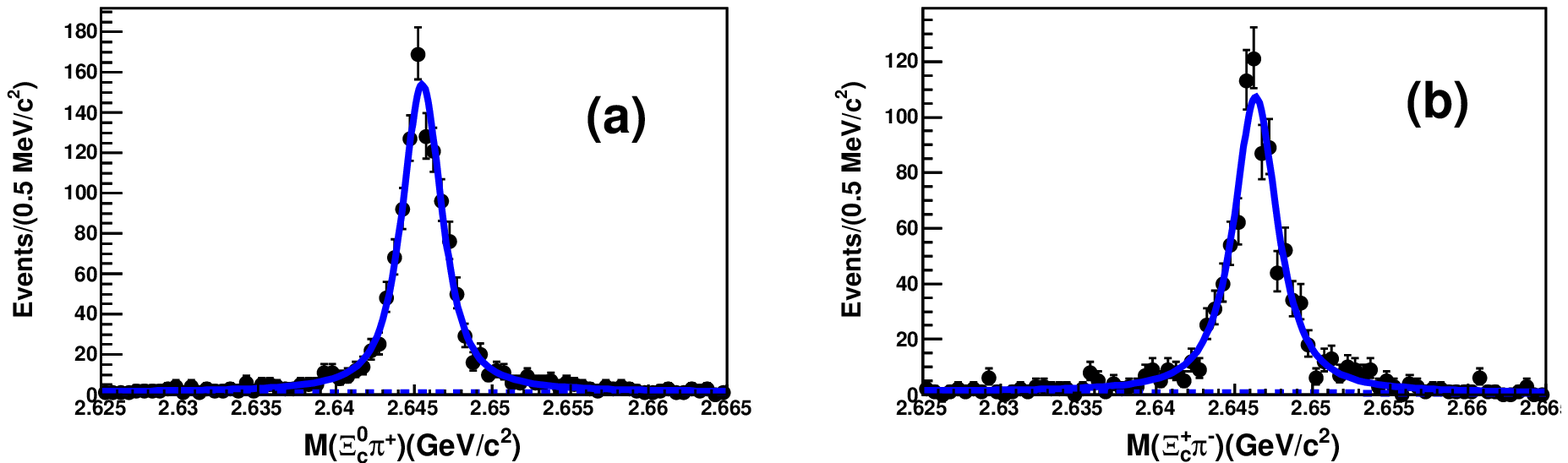}
\caption{ The a) $\Xi_c^0\pi^+$ and b) $\Xi_c^+\pi^-$ invariant mass distributions with a selection on the $\Xi_c\pi\pi$ 
invariant mass of the $\Xi_c(2815)$ parent state and a scaled momentum cut of $x_p > 0.7$ on the parent state. Both show clear $\Xi_c(2645)$ signals. The fits are described in the text. The dashed lines represent the combinatorial background
contributions. }
\label{fig:JAllData2645R}
\end{figure}

Relaxing the mass cut around the $\Xi_c(2645)$ peaks and instead selecting events in the $\Xi_c\pi\pi$ spectrum 
within 6 ${\rm MeV}/c^2$ of the $\Xi_c(2815)$ signal,
we can study the $\Xi_c^0\pi^+$ and $\Xi_c^+\pi^-$ mass spectra (Fig.~\ref{fig:JAllData2645R}). 
Using this method of selection, rather than looking at these combinations inclusively,
produces $\Xi_c(2645)$ signals of much higher signal-to-noise ratio than is possible in inclusive studies~\cite{KATO2}, 
and reduces the dependence on the shape of the background,
which is especially important as any background may include ``satellite'' peaks from partially reconstructed resonances.
The two distributions are fit, as in the case of the $\Xi_c(2815)$, to obtain masses of 
$(2645.44\pm0.06)$ ${\rm MeV}/c^2$ for the 
$\Xi_c^+(2645)$ and $(2646.32\pm 0.07)$ ${\rm MeV}/c^2$ 
for the $\Xi_c^0(2645)$, with intrinsic widths of $\Gamma(\Xi_c^+(2645))=(2.04\pm0.14)$ ${\rm MeV}$ and 
$\Gamma(\Xi_c^0(2645))=(2.26\pm0.18)$ ${\rm MeV}$, where the uncertainties quoted are statistical only, and the systematic uncertainties 
discussed in section on systematic uncertainties. We note that the requirements of 5 ${\rm MeV}/c^2$ and 6 ${\rm MeV}/c^2$ 
detailed above are sufficiently loose to not significantly bias the subsequent 
measurements.

\section*{The \boldmath${\Xi_c(2980) \to \Xi_c(2645)\pi}$ decay}

\begin{figure}[htb]
\includegraphics[width=7.in]{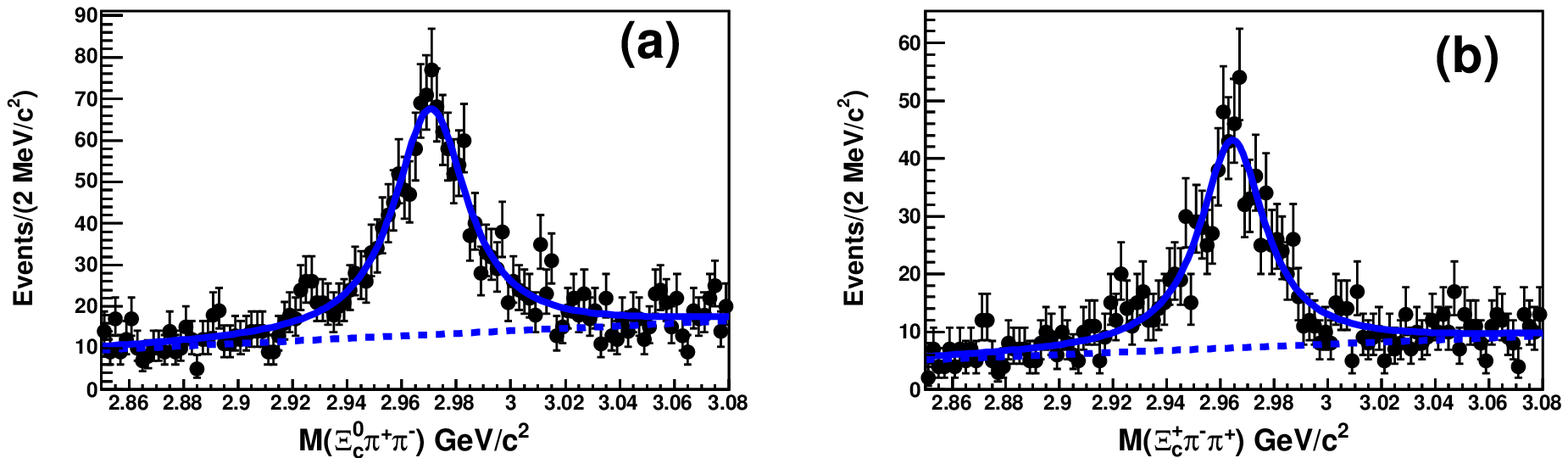}

\caption{ The a) $\Xi_c^0\pi\pi$ and b) $\Xi_c^+\pi\pi$ invariant mass distributions with a cut on $\Xi_c(2645)$ invariant mass of the intermediate state and a scaled momentum cut 
of $x_p > 0.7$. Both show clear $\Xi_c(2980) $ signals. The fits 
are described in the text. The dashed lines represent the 
combinatorial background contributions. }

\label{fig:JAllData2980}

\end{figure}

The $\Xi_c(2980)$ state was discovered in $\Xi_c(2645)\pi$ decay by Belle in 2005~\cite{CHISTOV}, and then found to also decay to 
$\Lambda_c^+ K^-\pi^+$~\cite{LESIAK}.
The analysis of this state is identical to that of the $\Xi_c(2815)$ above except that 
we focus on the mass range 2.84-3.10 ${\rm GeV}/c^2$ and 
present the 
mass distributions (Fig.~\ref{fig:JAllData2980}) in bins appropriate for the width of the signal. We choose not to try to fit the mass range extending from 
below the $\Xi_c(2815)$ to the $\Xi_c(2980)$ with one fit, as the background shape may include undulations due to combinations of partially
reconstructed excited $\Xi_c$ states with other pions. 

As in the previous case, the signal shape is a Breit-Wigner convolved with a double-Gaussian resolution function. This gives signal yields as shown 
in Table~\ref{tab:yields}. The measured masses and widths are $M(\Xi_c^0(2980)) = (2970.8\pm0.7)$ ${\rm MeV}/c^2$, $M(\Xi_c^+)=(2966.0\pm0.8 )$ ${\rm MeV}/c^2$, 
$\Gamma(\Xi_c^0(2980))=(30.3\pm2.3)$ ${\rm MeV}$,
$\Gamma(\Xi_c^+(2980))=(28.1\pm2.6)$ ${\rm MeV}$, 
where the uncertainties shown are purely statistical.

\section*{The \boldmath${\Xi_c^{\prime} \to \Xi_c}$ decay}

The $\Xi_c^{\prime}$ doublet is the charmed-strange analog of the $\Sigma_c(2455)$ triplet. However, because their mass difference 
with respect to the ground state
is less than the pion mass, the $\Xi_c^{\prime}$ particles are found by their electromagnetic decays, 
$\Xi_c^{\prime} \to \Xi_c\gamma$, and their intrinsic
widths are experimentally negligible. 
The photons, detected in the electromagnetic calorimeter, are required to have an energy greater than 100 ${\rm MeV}$ in the laboratory
frame, and to have a transverse shape consistent with that expected for a single photon. Each photon is combined with the 
$\Xi_c$ candidates, as detailed
above, and the $\Xi_c\gamma$ mass distribution is plotted for combinations with a value of $x_p > 0.65$.
This lower choice of the $x_p$ requirement is because the $\Xi_c^{\prime}$ is not orbitally excited and is often a decay product
of higher-mass states. 
The shape of the background is complicated by the fact that
the low energy threshold allows many fake photons, not necessarily emanating from the collision, to be included. 
There are clearly many $\pi^0$ transitions
from excited $\Xi_c$ states that produce photons correlated with the ground-state $\Xi_c$ baryons.
It is not possible to improve the signal and background separation by vetoing 
photons that can be combined with another photon to make a $\pi^0$, 
as the number of fake photons is too high. 

\begin{figure}[htb]																	 
\includegraphics[width=7in]{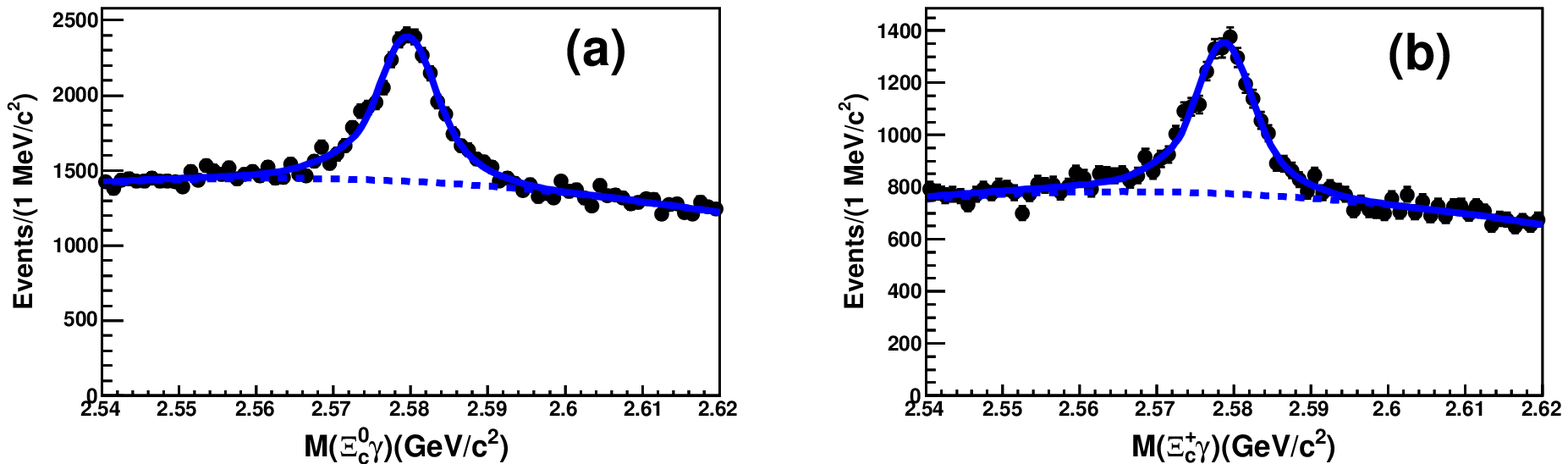}
\caption{ The a) $\Xi_c^0\gamma$ and b) $\Xi_c^+\gamma$ invariant mass distributions. The fits are described in the text. 
The dashed lines represent the combinatorial
background contributions. }
\label{fig:Xicprime}
\end{figure}

The $\Xi_c\gamma$ mass distributions (Fig.~\ref{fig:Xicprime}) 
are fit to the sum of a polynomial background function and two ``Crystal Ball''~\cite{CB} 
functions to parameterize the signal. The Crystal Ball function is a Gaussian with an exponential tail on the low mass end. 
In this case we add two Crystal Ball functions in a fixed ratio with differing Gaussian resolution components but
all other parameters, derived using Monte Carlo modeling, at fixed values.
The asymmetry of the functional form naturally leads to a mass shift, whereby the most likely mass is not the same as the peak mass found
from the fit. This effect is modeled in Monte Carlo, and results in a 0.36 ${\rm MeV}/c^2$ shift. 
The measured masses, taking into account this shift, are $(2579.2\pm0.1)$ ${\rm MeV}/c^2$ and 
$(2578.4\pm0.1)$ ${\rm MeV}/c^2$ for the	$\Xi_c^{\prime^+}$ and $\Xi_c^{\prime^0}$, 
respectively, where the uncertainties
are statistical only; the systematic uncertainties are discussed below.

\section*{ Study of \boldmath$\Xi_c^\prime \pi$ combinations}

The $\Xi_c^\prime$ candidates detailed above are mass-constrained to their measured mass values, 
and then the combinations of these candidates with appropriately charged pions in the events are made to search 
for excited resonances decaying to $\Xi_c^{\prime 0}\pi^+$ or $\Xi_c^{\prime +}\pi^-$.

\begin{figure}[htb]														      
\includegraphics[width=7in]{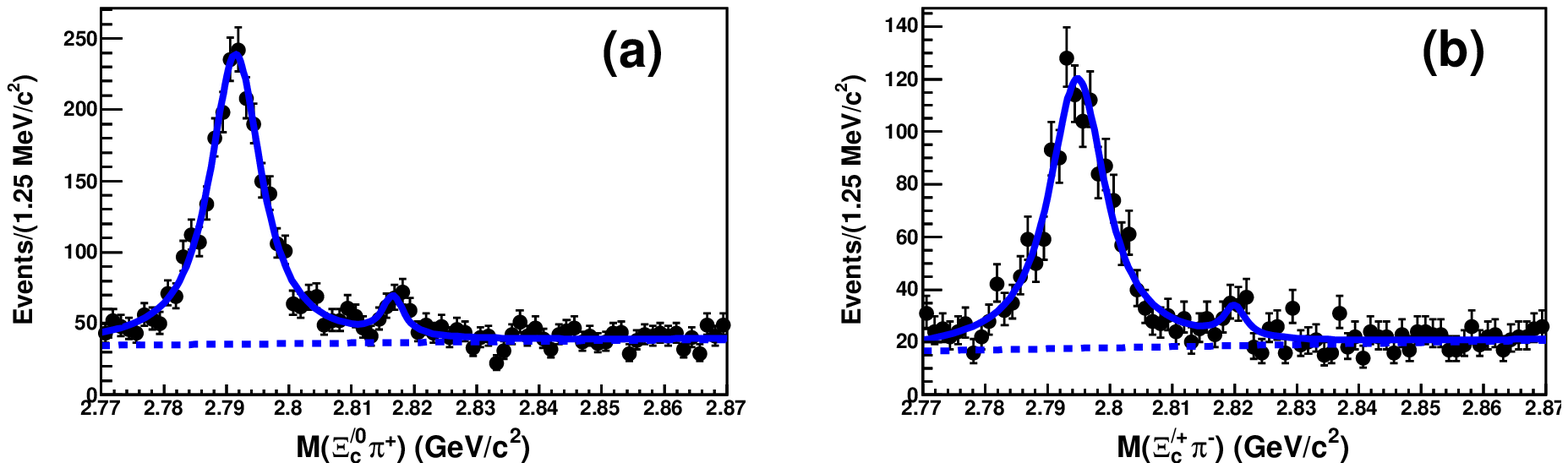}
\caption{The a) $\Xi_c^{\prime 0}\pi$ and b) $\Xi_c^{\prime +}\pi$ invariant mass distributions. The fits
are described in the text. The dashed lines represent the combinatorial background contributions.}
\label{fig:Fig2790}
\end{figure}

The mass distributions for these combinations, with a requirement of $x_p > 0.7$, are shown in Fig.~\ref{fig:Fig2790}, and both show large
$\Xi_c(2790)$ peaks and smaller peaks in the $\Xi_c(2815)$ region. 
The low-mass cut-offs of the mass ranges were chosen to
exclude the satellite peaks found from fake $\Xi_c^{\prime}$ combinations with a transition pion from a $\Xi_c(2645)\to\Xi_c$ decay. 
It is possible for background photons, particularly of low energy, to combine with the $\Xi_c$ ground states to make $\Xi_c^{\prime}$ candidates. 
Once constrained to the $\Xi_c^{\prime}$ mass, several such candidates in one event can combine with a pion from a higher state to make
multiple entries in this plot, all at similar total masses. To avoid this, we require that if there are multiple $\Xi_c^{\prime}$ candidates
of this type in an event, only the one with an unconstrained mass closest to the $\Xi_c^{\prime}$ mass is considered. This reduces the overall population of the
plot by around 15\%. 

Each distribution is fit to the sum of a polynomial background function, and two signal shapes. 
The signal shapes are each Breit-Wigners convolved with a double-Gaussian resolution 
function (as shown in Table~\ref{tab:MC}) to parameterize the $\Xi_c(2790)\to\Xi_c^{\prime}$ 
and $\Xi_c(2815)\to\Xi_c^{\prime}\pi$ decays. 
The masses and intrinsic widths of the $\Xi_c(2815)$ states are fixed to those found in the analysis detailed above. 

The masses and the widths of the $\Xi_c(2790)$ states, with appropriate corrections as described above, are 
found to be $M(\Xi_c^+(2790))=(2791.6\pm0.2$) ${\rm MeV}/c^2$ and $\Gamma(\Xi_c^+(2790)) = (8.9\pm0.6\pm0.8) $ ${\rm MeV}$, 
and $M(\Xi_c^0(2790))=(2794.9\pm0.3)$ ${\rm MeV}/c^2$ and 
$\Gamma(\Xi_c^0(2790)) = (10.0\pm0.7\pm0.8)$ ${\rm MeV}$. 
This is the first observation of significantly non-zero widths for these particles, 
although the original CLEO paper~\cite{half} indicated that it was likely that they
had intrinsic widths of this order. The estimation of the systematic uncertainties on the masses 
follows the method used for the other states under investigation, and is described below. 
In this case, there is a systematic uncertainty 
due to the uncertainty in the M($\Xi_c^{\prime})-M(\Xi_c)$ mass difference, as well as the uncertainties in the masses
of the ground states. 

\begin{figure}[htb]														      
\includegraphics[width=7in]{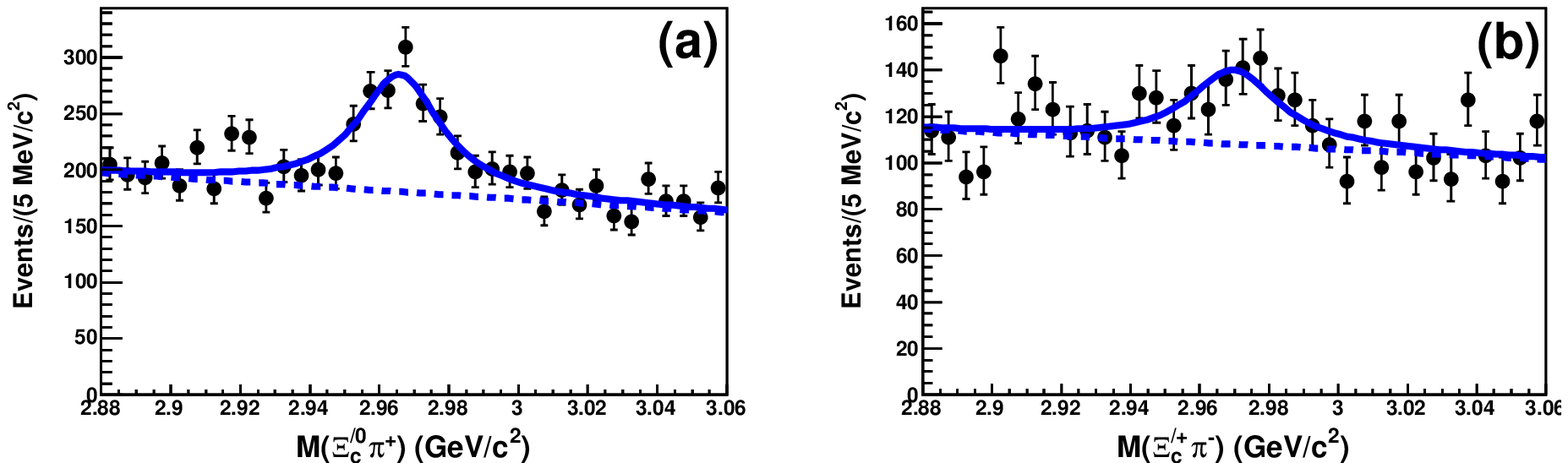}
\caption{ 
The  a) $\Xi_c^{\prime 0}\pi^+$ and b) $\Xi_c^{\prime +}\pi^-$ invariant mass distributions. The fits
are described in the text. The dashed lines represent the combinatorial background contributions.}

\label{fig:Fig2970}
\end{figure}

Yields of $128\pm25$  $\Xi_c^+(2815)\to\Xi_c^{\prime 0}\pi^+$ events and
$52\pm17$ $\Xi_c^0(2815)\to\Xi_c^{\prime +}\pi^-$, events are obtained.
The central values of the corresponding mass peaks have an uncertainty 
as we are using the mass measurement via a decay chain into the neutral 
ground state to study the decay into the charged ground state, and vice versa.   

It is not possible, from this study, to extract the branching ratio 
$B(\Xi_c(2815)\to\Xi_c^\prime\pi) /B(\Xi_c(2815)\to\Xi_c(2645)\pi)$
as we do not know the relative production cross 
sections of the ground-state $\Xi_c$ baryons, or their
absolute branching fraction in any one mode. 
However, we can perform a ``back-of-the-envelope'' calculation on the assumption that the efficiency times the 
branching fractions for the reconstructed modes of the two ground states indicated are equal to the ratio of those reconstructed,
which is equivalent to assuming that the ground-state $\Xi_c^0$ and $\Xi_c^+$ are produced in equal numbers. 
Using this, we can infer that $B(\Xi_c^+(2815)\to\Xi_c^{\prime 0}\pi^+) /B(\Xi_c^+(2815)\to\Xi_c^0(2645)\pi^+,\Xi_c^0(2645)\to\Xi_c^+\pi^-)\approx 11\%$,
and $B(\Xi_c^0(2815)\to\Xi_c^{\prime +}\pi^-) /B(\Xi_c^0(2815)\to\Xi_c^+(2645)\pi^-,\Xi_c^+(2645)\to\Xi_c^0\pi^+)\approx 10\%$.
In each of these branching fractions, the denominator and numerator involve different matrix elements~\cite{HQS}.

\section*{Observation of \boldmath$\Xi_c(2980) \to  \Xi_c^{\prime}\pi$ decays}

Lastly, we search for decays of the type $\Xi_c(2980)\to\Xi_c^{\prime}\pi$. Figure~\ref{fig:Fig2970} shows the same distributions as
Fig.~\ref{fig:Fig2790}, plotted in a different mass region.
We fit each distribution to the sum of a linear background function and signal functions using masses and intrinsic widths
from our measurements above, and convolved with double-Gaussian resolution functions.
The fitted yields are $845\pm77$ and $276\pm59$, for the charged and neutral parent states, respectively. 

Once again, we cannot accurately measure the relative branching fractions, but can estimate
 that $B(\Xi_c^+(2980)\to\Xi_c^{\prime 0}\pi^+) /B(\Xi_c^+(2815)\to\Xi_c^0(2645)\pi^+,\Xi_c^0(2645)\to\Xi_c^+\pi^-)\approx 75\%$,
and $B(\Xi_c^0(2980)\to\Xi_c^{\prime +}\pi^- )/B(\Xi_c^0(2815)\to\Xi_c^+(2645)\pi^-,\Xi_c^+(2645)\to\Xi_c^0\pi^+)\approx 50\%$.
The apparently large branching fraction of the $\Xi_c(2980)$ into $\Xi_c^{\prime}$ may prove useful in identifying the nature of the state. It does
appear to be consistent with the decays of the $\Lambda_c(2765)$\cite{ARTUSO,JOO}, which is of similar excitation energy above its ground state. 
One possible interpretation is
that they are radial excitations of the ground-state charmed baryons~\cite{EBERT}.

\section*{Systematic Uncertainties}

In general, the resolution of the masses of resonances decaying via transition pions is dominated 
by the resolution of the momentum measurements of those pions rather than that of the ground states.
However, Monte Carlo simulation predicts that decays to the different decay modes of the ground-state
$\Xi_c$ and $\Xi_c^+$ can have resolutions varying by $\approx\pm 10\%$ of the average value. 
Therefore, in finding the resolution functions to be used, care is taken to generate the various
decay modes in the correct proportions to correctly reproduce the fractions found in the data. 
The events were also generated using the same range of beam
energies as the real data, and with a fragmentation function (\emph{i.e.} $x_p$ distribution) that matches the data, as the resolution does vary 
slowly with momentum
(with a change of $\approx 10\% $ for the full momentum range under consideration). 

A variety of checks are performed to ensure that the Monte Carlo generation technique was a good representation of the reconstructed data.
For the final result on the masses of the particles, we separate the systematic uncertainty of the measurement of the mass differences, 
which is in effect what we are measuring, from that of the masses of the ground states tabulated by the Particle Data Group~\cite{PDG}. 
The overall momentum scale of the Belle detector is 
well studied from checks with $K^0_S$ and other particle masses accurately known from other experiments.
However, detailed studies of the $D^{*+}\to D^0 \pi^+$ decay indicates that there are deviations particularly associated with the 
reconstruction and fitting of low-momentum charged tracks. As has been described above, this does lead to small systematic offsets
in the found masses, and so the results are corrected for these effects. The uncertainty in these corrections is conservatively set
at 50\% of the offset, and these are listed as the uncertainty due to the mass scale in Table~\ref{tab:systematics}.
  
The resolution in the Monte Carlo simulation matches 
the measured resolution within $8\%$ for all the ground-state $\Xi_c$ decay modes presented here. 
Further tests, specifically on the measurement of 
mass differences, were performed using the $D^{*+}\to D^0 \pi^+$ and the $\Lambda_c(2625)\to\Lambda_c^+\pi^+\pi^-$ decays. 
The latter uses an analysis chain very similar to the ones under consideration in this analysis, and, 
by matching the signal in the data with that in Monte Carlo simulated data with zero intrinsic width, 
we can conclude that the calculated width of the resolution cannot be more than $5\%$ too broad for this topology. 
After assessment of the results of all of the above and other checks on the resolution, we assign a $10\%$ uncertainty to the width of the resolution 
function, and assign the systematic uncertainties on the intrinsic widths accordingly.

All signal shapes used to analyze the strong decay transitions are Breit-Wigner functions convolved with double-Gaussian resolution 
functions. Technically, this is performed using two ``Voigtian'' functions in the RooFit fitting package~\cite{RooFit}. 
Likelihood fits were performed using a large
number of small bins - but are presented with bin sizes appropriate for each individual plot - so that there are negligible uncertainties associated 
with binning. Fits are also performed using relativistic Breit-Wigner 
functions that include spin-dependent and mass-dependent widths. As the widths of the particles are small compared with the $Q^2$ associated with the
decays, the extracted masses and widths do not depend greatly on the choice of signal functions. However, the 
small differences were considered as the systematic uncertainties associated with the signal shape. We do not use the relativistic
Breit-Wigner functions for the default fits, as the spin-parity of the states have not been determined and, in addition, there are complications 
in the phase-space description of the decays when the decay product is itself of non-neglible intrinsic width.

The combinatorial backgrounds are parameterized by low-order Chebyshev polynomials. The small variations found by varying the order of this
polynomial by one are taken as the systematic uncertainty due to the choice of background function. There is a specific issue with the background
parameterization for the $\Xi_c(2980)$ mass distribution of Fig.~\ref{fig:JAllData2980}. Here, there is some evidence for peaking at around 
$M=2.92$ ${\rm GeV}/c^2$,
particularly in the neutral state.
This evidence is not sufficient to claim the existence of a new particle, and may be 
an unidentified ``satellite'' peak formed from 
partially reconstructed resonances. Allowing an additional signal function to appear near this value changes
the extracted width and mass values, lowering the extracted width measurement. 
This is particularly the case in the charged state, even though the evidence of a non-polynomial background shape is 
less convincing. 
These possible changes in values were considered part of the systematic uncertainty due to the background parameterization. 

One of many checks on the photon energy scale, and the corrections made because of the asymmetric line-shape, was the reconstruction of 
$D^{*0}\to D^0\gamma$ using a very similar analysis. The measured mass difference was found to be $0.15\pm0.10$ MeV higher than the Particle
Data Group value~\cite{PDG}, which is dominated by measurements using the low $Q^2$ decay $D^{*0}\to D^0\pi^0$.
The assigned systematic uncertainty of 0.4 MeV for the $\Xi_c^{/+}$ and $\Xi_c^{/0}$ mass differences with respect to the ground
state is greater than this small discrepency.

In Table~\ref{tab:systematics}, we summarize the systematic uncertainties for the mass and width measurements of the five isodoublets under study. 
With the exception of the particular case of the background shape for the $\Xi_c(2980)$ discussed above, 
if the method used to evaluate these uncertainties yields slightly different values for the two charged states, the greater of the two is used.

\begin{table}[htb]

\caption{The systematic uncertainties for the mass (in ${\rm MeV}/c^2$) and width measurements (in ${\rm MeV}$). 
``Mass scale'' refers to the uncertainty in making a mass measurement
in the particular kinematic region under investigation. ``Resolution'' refers to uncertainty in the 
Monte Carlo correctly modeling the mass resolution,
``Signal shape'' refers to which version of a Breit-Wigner shape is used, and ``Background Shape'' 
refers to the uncertainty due to the use of different
acceptable formulisms of the combinatorial background. The total systematic uncertainty is the quadratic sum of the individual contributions.}
\begin{tabular}
%{@{\hspace{0.5cm}}l@{\hspace{0.5cm}}||@{\hspace{0.5cm}}c@{\hspace{0.5cm}}}
 {@{\hspace{0.5cm}}l@{\hspace{0.5cm}}|  
@{\hspace{0.5cm}}c@{\hspace{0.5cm}}|
@{\hspace{0.5cm}}c@{\hspace{0.5cm}}|
@{\hspace{0.5cm}}c@{\hspace{0.5cm}}| 
@{\hspace{0.5cm}}c@{\hspace{0.5cm}}| 
@{\hspace{0.5cm}}c@{\hspace{0.5cm}} 
  }
\hline \hline
  & Mass scale  & Resolution & Signal shape & Background shape & Total \\
\hline

$M(\Xi_c(2645))$  & 0.04  & 0.0 & 0.06 & 0.01 & 0.07   \\
$M(\Xi_c(2815))$  & 0.06  & 0.0 & 0.03 & 0.01 & 0.07   \\
$M(\Xi_c(2980))$  & 0.14  & 0.0 & 0.12 & 0.05 & 0.2   \\
$M(\Xi_c^{\prime})$  & 0.4  & 0.0 & 0.03 & 0.03 & 0.4   \\
$M(\Xi_c(2790))$  & 0.06  & 0.0 & 0.04 & 0.05 & 0.1   \\

\hline

$\Gamma(\Xi_c(2645))$  & 0.0  & 0.10 & 0.06 & 0.03 & 0.13   \\
$\Gamma(\Xi_c(2815))$  & 0.0  & 0.16 & 0.02 & 0.10 & 0.17   \\
$\Gamma(\Xi_c(2980)^0)$  & 0.0  & 0.1 & 0.7 & $+0.7,-1.8$ & $+1.0,-1.8$   \\
$\Gamma(\Xi_c(2980)^+)$  & 0.0  & 0.1 & 0.7 & $+0.7,-5.0$ & $+1.0,-5.0$   \\
$\Gamma(\Xi_c(2790))$  & 0.0  & 0.2 & 0.3 &  0.6 & 0.8    \\
\hline
\hline

\end{tabular}
\label{tab:systematics}

\end{table}

\section*{Results}

Table~\ref{tab:results} shows the results of the measurements of the masses and widths of the five isodoublets. In all cases, 
the first uncertainty is statistical, and the second is the systematic uncertainty associated with the individual measurement. All the masses have a final,
asymmetric uncertainty, taken from the Particle Data Group~\cite{PDG}, for the mass of the ground states, and the $\Xi_c(2790)$ have an extra uncertainty due
to the uncertainty in the $M(\Xi_c^{\prime})-M(\Xi_c)$ measurement. The results are presented in this manner so that the final masses may be adjusted
should new measurements on the ground states become available.

\begin{table}[htb]
\caption{The final results for the masses (in ${\rm MeV}/c^2$) and widths (in MeV) for the five isodoublets under study. For comparison, the
2015 world averages~\cite{PDG} (denoted ``PDG'') are also quoted. Mass differences are with respect
to the daughter states.}
\label{tab:results}

\begin{tabular}
{
@{\hspace{0.1cm}}l@{\hspace{0.1cm}}|
@{\hspace{0.1cm}}c@{\hspace{0.1cm}}|
@{\hspace{0.1cm}}c@{\hspace{0.1cm}}|  
@{\hspace{0.1cm}}c@{\hspace{0.1cm}}|
@{\hspace{0.1cm}}c@{\hspace{0.1cm}}|
@{\hspace{0.1cm}}c@{\hspace{0.1cm}}  
 }
\hline \hline
Particle  & Yield & Mass   & $M-M(\Xi_c)$ & $M-M(\Xi_c^{\prime}$) & Width  \\
\hline

 $\Xi_c(2645)^+$  & $1260\pm40$  & $2645.58\pm 0.06\pm 0.07^{+0.28}_{-0.40} $ & $174.66\pm 0.06\pm 0.07$ &   &  $2.06\pm 0.13\pm 0.13$   \\
 PDG              &       &$2645.9\pm 0.5$                                   &   $175.0\pm0.6$          &   &  $2.6\pm 0.2 \pm 0.4$ \\
$\Xi_c(2645)^0$  & $975\pm36$  & $2646.43\pm 0.07\pm 0.07^{+0.28}_{-0.40} $ & $178.46\pm 0.07\pm 0.07$ &   &  $2.35\pm0.18\pm0.13$   \\
PDG              &       &$2645.9\pm 0.5$                                   &      $178.0\pm0.6$       &   &  $< 5.5 $ \\
\hline
$\Xi_c(2815)^+$  & $941\pm 35$  & $2816.73\pm 0.08\pm 0.06^{+0.28}_{-0.40}$ & $348.80\pm0.08\pm0.06$ &  & $2.43\pm0.20\pm0.17$   \\
PDG               &      &$2816.6\pm 0.9$                         & $348.7\pm0.9$                          &   &  $<3.5$ \\
$\Xi_c(2815)^0$  & $1258\pm 40$  & $2820.20\pm 0.08\pm 0.07^{+0.28}_{-0.40}$  & $349.35\pm 0.08\pm 0.07$ & & $2.54\pm 0.18 \pm 0.17$ \\                  
PDG               &      &$2819.6\pm 1.2$                         &     $348.8\pm1.2$                     &   &  $<6.5$ \\
\hline
$\Xi_c(2980)^+$  & $916\pm 55$  &  $2966.0\pm 0.8\pm 0.2^{+0.3}_{-0.4}$  & $498.1 \pm 0.8 \pm 0.2$ &  & $28.1\pm 2.4 ^{+1.0}_{-5.0} $   \\
PDG               &      &$2970.7 \pm 2.2 $                         &                          &   &  $17.9 \pm 3.5$  \\
$\Xi_c(2980)^0$  & $1443\pm 75$  & $2970.8\pm 0.7 \pm 0.2^{+0.3}_{-0.4}$ & $499.9\pm 0.7 \pm 0.2$ &  & $30.3\pm 2.3 ^{+1.0}_{-1.8}$  \\
PDG               &      &$2968.0\pm 2.6\pm 0.5$                         &            &   &  $20\pm 7$ \\
\hline
$\Xi_c^{\prime +}$  & $7055\pm211$  & $2578.4\pm 0.1 \pm 0.4^{+0.3}_{-0.4}$ & $110.5\pm 0.1 \pm 0.4$ &  & \\  
PDG               &      &$2575.6\pm 3.0 $                         &  $107.8\pm 3.0$   &   &   \\
$\Xi_c^{\prime 0}$  & $11560\pm276$  & $2579.2\pm 0.1 \pm 0.4^{+0.3}_{-0.4}$ & $108.3\pm0.1\pm0.4$  & &  \\
PDG               &      &$2577.9\pm 2.9$                         &   $107.0\pm 2.9$     &   &   \\
\hline
$\Xi_c(2790)^+$  & $2231\pm103$  & $2791.6\pm 0.2 \pm 0.1 \pm 0.4^{+0.3}_{-0.4}$ & $320.7\pm0.2\pm0.1\pm 0.4$ & $213.2\pm0.2 \pm0.1$ & $8.9\pm0.6\pm0.8$  \\
PDG               &      &$2789.8\pm 3.2$                         &                     $318.2\pm3.2$     &   &  $<15$ \\
$\Xi_c(2790)^0$  & $1241\pm72$  & $2794.9\pm 0.3\pm 0.1 \pm 0.4^{+0.3}_{-0.4}$ & $323.8\pm0.2\pm 0.1\pm 0.4$ & $215.7\pm0.2\pm 0.1$ & $10.0\pm 0.7\pm 0.8$ \\
PDG               &      &$2791.9\pm3.3$                         &                  $324.0\pm3.3$        &   &  $<12$ \\

\hline
\hline
 
\end{tabular}

\end{table}

The systematic uncertainties associated with the mass scales cancel in the measurement of 
isospin splittings. However, in each case, there is an uncertainty that arises from the 
measurement of splitting of the ground states, $M(\Xi_c^+) - M(\Xi_c^0)$ = $(-2.92\pm0.48)$ ${\rm MeV}/c^2$.
This is reported separately from the statistical and systematic uncertainties associated with this 
measurement so that the total uncertainties may be reduced when new measurements of the masses of
the ground states become available.

\begin{table}[htb]
\caption{The isospin splitting between the members of each isodoublet.}

\label{tab:isospin}

\begin{tabular}
{
@{\hspace{0.1cm}}l@{\hspace{0.1cm}}|
@{\hspace{0.1cm}}c@{\hspace{0.1cm}}
  }
\hline \hline
Particle  &  $M(\Xi_c^+)-M(\Xi_c^0)$ (${\rm MeV}/c^2$)  \\
\hline

$\Xi_c(2645)$      & $-0.85\pm 0.09 \pm 0.08 \pm0.48$            \\
$\Xi_c(2815)$      & $ -3.47\pm 0.12 \pm 0.05 \pm 0.48$              \\
$\Xi_c(2980)$      & $-4.8\pm 0.1 \pm 0.2 \pm 0.5$             \\
$\Xi_c^{\prime}$  & $-0.8\pm 0.1\pm 0.1\pm 0.5$         \\
$\Xi_c(2790)$      & $-3.3\pm 0.4\pm 0.1\pm 0.5$          \\

\hline
\hline
 
\end{tabular}

\end{table}

\section*{Comparisons with Theoretical Models}

Of the five excited $\Xi_c$ isodoublets investigated here, there are clear spin-parity assignments for four of them. The exception 
is the copiously produced $\Xi_c(2980)$. The evidence presented here that the $\Xi_c(2980)$ states decay significantly 
to $\Xi_c^{\prime}\pi^+$ may be used to clarify the situation. We note that the $\Xi_c(2980)$ may be the strange analog of the
$\Lambda_c(2765)$, which also has a high cross section, and appears to decay to $\Sigma_c(2455)$ and $\Sigma_c(2520)$~\cite{ARTUSO,JOO}. 
A possible interpretation
of these states is that they represent radial excitations of the ground-state charmed baryons~\cite{EBERT}. 

Many models relate the intrinsic widths of the particles in the charmed baryon spectrum. The strange charmed sector and the non-strange charmed 
sector (\emph{i.e.} $\Lambda_c/\Sigma_c$ ) give complementary information on what, in HQS, are the same coupling constants. For instance, using the 
input from the measurement of the  $\Lambda_c^+(2595)$ width~\cite{PDG} leads to predictions of the intrinsic width of the $\Xi_c(2790)$ baryons.
However, those measurements are complicated by the fact that the  $\Lambda_c^+(2595)\to\Sigma_c\pi$ decay occurs very close to threshold, and
how exactly its mass is 
measured and the distortion of its line-shape is treated can greatly influence the predictions. For instance, Cheng and Chua~\cite{CC2},
using the latest values for the $\Lambda_c^+(2595)$, 
predict $\Gamma(\Xi_c^+(2790))=16.7^{+3.6}_{-3.6}$ ${\rm MeV}$ and $\Gamma(\Xi_c^0(2790))=17.7^{+2.9}_{-3.8}$ ${\rm MeV}$, 
whereas, using earlier values, they found  
$\Gamma(\Xi_c^+(2790)=8.0^{+4.7}_{-3.3}$ ${\rm MeV}$ and $\Gamma(\Xi_c^0(2790))=8.5^{+5.0}_{-3.5}$ ${\rm MeV}$, which are 
very close to our experimental measurements. 
The measurements of the $\Xi_c(2790)$ states allow far more robust measurements of the coupling constants to be made.

Similarly, the measurements of $\Xi_c(2815)$ widths are in reasonable agreement with some previous predictions~\cite{CC1,Tawfiq}, but less so 
with Cheng and Chua's latest model that predicts $\Gamma \approx 7.4\ {\rm MeV}$~\cite{CC2}. 
The analogous decay to $\Xi_c(2815)\to\Xi_c(2645)\pi$ in the $\Lambda_c/\Sigma_c$ 
sector is $\Lambda_c^+(2625)\to\Sigma_c(2520)\pi$ which is not kinematically allowed,
again complicating the extraction of the relevant parameters. 

The measurements of the widths of the $\Xi_c(2645)$ baryons are in good agreement with calculations. Here the predictions are more robust as
they rely upon the measurements of 
the $\Sigma_c^{++}$ and $\Sigma_c^0$ baryons, which
 have closely analagous decays. 
Using the recent high-precision values of $\Sigma_c$ widths~\cite{SIGC}, 
Cheng and Chua predict $\Gamma(\Xi_c^+(2645))=2.4^{+0.1}_{-0.2}$ ${\rm MeV}$ and
$\Gamma(\Xi_c^0(2645))=2.5^{+0.1}_{-0.2}$ MeV, close to our new measurements,
further validating the use of HQS in the charm sector.

The isospin splittings of charmed baryons are due to the difference in the $u$ and $d$ 
quark masses together with electromagnetic 
interactions. In 2003 several versions of non-relativistic quark model~\cite{SIL} predicted 
small ($<1$ ${\rm MeV}/c^2$) 
splittings in the $\Xi_c^{\prime}$ and $\Xi_c(2645)$ systems, but a splitting of $\approx 3 $ ${\rm MeV}/c^2$ 
in the $\Xi_c(2815)$ system, similar to the ground states and in good agreement with the results presented here.
The sizeable measured splittings of the 
$\Xi_c^+(2980)$ and $\Xi_c^0(2980)$ may help to identify the states. The $\Xi_c^{\prime}$ system has been the subject
of some theoretical interest, and several authors~\cite{GUO} predict
small negative isospin splittings, in agreement with our measurements.     

\section*{Summary and Conclusions}
 
Using the entire $980\ {\rm fb}^{-1}$ of data recorded by the Belle detector at the KEKB $e^+e^-$ collider operating in the $\Upsilon$ energy range,
we present new measurements of the masses of all members of five isodoublets of excited $\Xi_c$ states, and intrinsic widths of those
that decay strongly. Of the eighteen measurements, five are of intrinsic widths of particles for which only limits existed previously.
Of the remaining thirteen measurements, ten are within one standard deviation of the Particle Data Group~\cite{PDG} best-fit values.
The three measurements that are in modest disagreement with previous results are in the $\Xi_c(2980)$ sector, where the previous measurements
were dominated by decays into different final states and for which 
some measurements may have been prone to the existence of more than one resonance in the
region, or biases from threshold effects. Although some of the previous measurements were made by Belle, they are 
all essentially independent of those presented here. For instance, although the measurement of $\Gamma(\Xi_c^+(2645))$ was made with the same
dataset, it was made using only three decay modes of the $\Xi_c^0$ but without the $\Xi_c(2815)$ tag, 
and the limiting systematic uncertainties are from completely different sources, thus making the measurements effectively complementary.

The intrinsic width measurements of the $\Xi_c(2790)$ and $\Xi_c(2815)$ states present a consistent picture with the instrinsic widths that can be
related to measurements in the $\Lambda_c/\Sigma_c$ system and used to predict measurements in the $b$-hadron sector. The mass measurements constitute
a considerable improvement in precision on previous measurements, and allow further investigation of hadron mass models including isospin splittings.

We thank the KEKB group for the excellent operation of the
accelerator; the KEK cryogenics group for the efficient
operation of the solenoid; and the KEK computer group,
the National Institute of Informatics, and the 
PNNL/EMSL computing group for valuable computing
and SINET4 network support.  We acknowledge support from
the Ministry of Education, Culture, Sports, Science, and
Technology (MEXT) of Japan, the Japan Society for the 
Promotion of Science (JSPS), and the Tau-Lepton Physics 
Research Center of Nagoya University; 
the Australian Research Council;
Austrian Science Fund under Grant No.~P 22742-N16 and P 26794-N20;
the National Natural Science Foundation of China under Contracts 
No.~10575109, No.~10775142, No.~10875115, No.~11175187, No.~11475187
and No.~11575017;
the Chinese Academy of Science Center for Excellence in Particle Physics; 
the Ministry of Education, Youth and Sports of the Czech
Republic under Contract No.~LG14034;
the Carl Zeiss Foundation, the Deutsche Forschungsgemeinschaft, the
Excellence Cluster Universe, and the VolkswagenStiftung;
the Department of Science and Technology of India; 
the Istituto Nazionale di Fisica Nucleare of Italy; 
the WCU program of the Ministry of Education, National Research Foundation (NRF) 
of Korea Grants No.~2011-0029457,  No.~2012-0008143,  
No.~2012R1A1A2008330, No.~2013R1A1A3007772, No.~2014R1A2A2A01005286, 
No.~2014R1A2A2A01002734, No.~2015R1A2A2A01003280 , No. 2015H1A2A1033649;
the Basic Research Lab program under NRF Grant No.~KRF-2011-0020333,
Center for Korean J-PARC Users, No.~NRF-2013K1A3A7A06056592; 
the Brain Korea 21-Plus program and Radiation Science Research Institute;
the Polish Ministry of Science and Higher Education and 
the National Science Center;
the Ministry of Education and Science of the Russian Federation and
the Russian Foundation for Basic Research;
the Slovenian Research Agency;
Ikerbasque, Basque Foundation for Science and
the Euskal Herriko Unibertsitatea (UPV/EHU) under program UFI 11/55 (Spain);
the Swiss National Science Foundation; 
the Ministry of Education and the Ministry of Science and Technology of Taiwan;
and the U.S.\ Department of Energy and the National Science Foundation.
This work is supported by a Grant-in-Aid from MEXT for 
Science Research in a Priority Area (``New Development of 
Flavor Physics'') and from JSPS for Creative Scientific 
Research (``Evolution of Tau-lepton Physics'').


\begin{thebibliography}{99}
\bibitem{CC}
Throughout this paper, 
the inclusion of the charge-conjugate mode decay is implied
unless otherwise stated.
\end{thebibliography}
\end{document}